\documentclass[prb,twocolumn,groupedaddress]{revtex4-1}
\usepackage{graphicx}
\usepackage{amsmath}

\begin{document}

\title{$^{63}$Cu and $^{199}$Hg NMR study of HgBa$_{2}$CuO$_{4+\delta}$ single crystals}

\author{Damian Rybicki$^{1,2}$} 
\email[]{rybicki@physik.uni-leipzig.de} 
\author{J\"urgen Haase$^1$}
\author{Marc Lux$^1$}
\author{Michael Jurkutat$^1$}
\author{Martin Greven$^3$}
\author{Guichuan Yu$^3$}
\author{Yuan Li$^4$}
\author{Xudong Zhao$^{3,5}$}
\affiliation{$^1$ Faculty of Physics and Earth Science, University of Leipzig, Linn$\acute{e}$stra\ss e 5, 04103 Leipzig, Germany}
\affiliation{$^2$ Faculty of Physics and Applied Computer Science, AGH University of Science and Technology, Al. Mickiewicza 30, 30-059 Krak\'{o}w, Poland}
\affiliation{$^3$ School of Physics and Astronomy, University of Minnesota, Minneapolis, Minnesota 55455, USA}
\affiliation{$^4$International Center for Quantum Materials, School of Physics, Peking University, Beijing 100871, China}
\affiliation{$^5$College of Chemistry, Jilin University, Changchun 130012, China}


\date{\today}

\begin{abstract}
We report on the temperature dependence of $^{63}$Cu and $^{199}$Hg NMR magnetic shifts and linewidths for an optimally doped and an underdoped HgBa$_{2}$CuO$_{4+\delta}$ single crystal, as well as the quadrupole splitting and its distribution for $^{63}$Cu. From the $^{63}$Cu and $^{199}$Hg \textit{magnetic shifts} we have recently concluded on the existence of two spin components with different temperature dependencies [J. Haase, D. Rybicki, C. P. Slichter, M. Greven, G. Yu, Y. Li, and X. Zhao, Phys. Rev. B 85, 104517 (2012)]. Here we give a comprehensive account of all data and focus on the linewidths and quadrupole splittings. While the $^{63}$Cu quadrupole coupling and its distribution are by and large temperature independent, we identify three regions in temperature for which the magnetic widths differ significantly: at the lowest temperatures the magnetic linewidths are dominated by the rigid fluxoid lattice that seems to have disappeared above about 60 K. In the intermediate temperature region, starting above 60 K, the magnetic linewidth is dominated by the spatial distribution of the magnetic shift due to the pseudogap spin component, and grows linearly with the total shift up to about \textit{$\sim$}170-230 K, depending on sample and nucleus. Above this temperature the third region begins with an sudden narrowing where the second, Fermi-liquid-like spin component becomes homogeneous. We show that all linewidths, quadrupolar as well as magnetic, above the fluxoid dominated region can be understood with a simple model that assumes a coherent charge density variation with concomitant variations of the two spin components. In addition, we find a temperature independent spin based broadening in both samples that is incoherent with the other broadening for the underdoped crystal, but becomes coherent for the optimally doped crystal.
\end{abstract}

\pacs{74.25.nj, 74.72.Gh, 74.72.Kf}


\keywords{Superconductivity, NMR, Pseudo-Gap}

\maketitle



\section{Introduction}
High-temperature superconductivity in the cuprates was discovered about a quarter century ago, but a full understanding of its origin is still lacking. An important property of these materials is their electron spin susceptibility, and its energy and wavevector dependence as a function of temperature ($T$) and hole concentration ($x$) holds important clues for theory. The \textit{uniform} magnetic response to an external, homogeneous magnetic field ($B_0$) is of particular importance and contains, in addition to spin also orbital contributions. First, there is the anisotropic, $T$ independent van-Vleck term from the bonding electrons. Secondly, the isotropic diamagnetic contribution from the core electrons (also $T$ independent). Thirdly, below the superconducting transition temperature ($T_{\rm c}$) we expect a small $T$-dependent screening of the field due to the superfluid in the mixed state (for large $B_0$). These contributions complicate the isolation of the spin part of the uniform susceptibility in magnetization or nuclear magnetic resonance (NMR) shift measurements. On the other hand, the uniform spin susceptibility is a simple, fundamental property of the cuprates and NMR is capable of testing by investigating different nuclear sites whether a single $T$-dependent spin component is at work or if one has to resort to more complicated scenarios. Given the structure of the doped cuprates with Cu and O holes, early NMR experiments addressed this question with the Cu and O studies on two underdoped cuprates\cite{Takigawa1991, Bankay1994} and it was concluded that a single $T$-dependent spin component is indeed appropriate for the understanding of these materials.  

More recently, in a careful study of La$_{1.85}$Sr$_{0.15}$CuO$_4$ it was found that the $T$ dependence of Cu and O shifts in this compound cannot be explained with a single $T$-dependent spin component.\cite{Haase2009} The authors showed that their data above $T_{\rm c}$ are in agreement with the scaling behavior suggested earlier from magnetization measurements that suggest a two-component approach.\cite{Johnston1989, Oda1990, Nakano1994}  
Such shift measurements are of great importance since they have significant consequences for the interpretation of other NMR parameters, as well. For example, the $T$ dependence of the nuclear relaxation rate that can be sensitive to very different wave vectors can lead to very different conclusions in a one- vs. a two-component scenario.\cite{Walstedt2011} 

In order to find out whether this two-component behavior is a peculiarity of La$_{2-x}$Sr$_x$CuO$_4$ or if it applies to other cuprates as well, we decided to study another single-layer system, HgBa$_{2}$CuO$_{4+\delta}$, for which high-quality single crystals have recently become available.\cite{zhao2006} We note that the two compounds, YBa$_2$Cu$_3$O$_{6.63}$ and YBa$_2$Cu$_4$O$_8$, from which NMR single-component behavior of the cuprates was deduced, are underdoped systems that have double CuO$_2$ layers and a lower structural symmetry.\cite{Takigawa1991, Bankay1994}  

HgBa$_{2}$CuO$_{4+\delta}$ is a perfect material for studying the spin susceptibility. Unlike La$_{1.85}$Sr$_{0.15}$CuO$_4$ it has one of the highest $T_{\rm c}$s up to date and has a very simple tetragonal crystal structure. There is only one CuO$_2$ plane and only one Cu site in the unit cell.\cite{Wagner1993, Antipov2002} The Cu sites in adjacent planes are linked by O-Hg-O, i.e., the apical oxygen atoms of Cu are connected through Hg. 
By increasing the content ($\delta$) of extra oxygen, which is located in the Hg plane, the hole concentration ($x$) in the CuO$_2$ plane increases. While the relation between $\delta$ and $x$ is not known exactly, $T_{\rm c}$ shows an approximately parabolic dependence on both parameters.\cite{Presland1991, Balagurov1999, Barisic2008} 
Since extra oxygen is located far away from the CuO$_2$ plane one may expect only a small influence on the structure of the CuO$_2$ plane. Single crystals of sufficient size were not available until recently, so that previous NMR measurements\cite{Itoh1996, Itoh1998, Gippius1997, Gippius1999, Suh1994, Suh1996, Suh1996a, Stern2005} could only be performed on (aligned) powders. Given the large shift anisotropies for the various nuclei and the large electric quadrupole interaction for Cu and O in the high-temperature superconducting cuprates (HTSC), single crystals are helpful for precise shift and linewidth analyses. Therefore, the recent success by some of us in synthesizing large, high purity single crystals of the single-layer HgBa$_{2}$CuO$_{4+\delta}$ \cite{zhao2006} stimulated their investigation also with NMR.

Here, we present a comprehensive set of Cu and Hg NMR shifts and linewidths data for an underdoped ($T_{\rm{c}}$=74 K) and an optimally doped ($T_{\rm{c}}$=97 K) sample of HgBa$_{2}$CuO$_{4+\delta}$. In separate accounts\cite{Rybicki2012, Haase2012} we found that the \textit{spin shifts} cannot be explained with a single $T$-dependent spin component, but with a two-component scenario where the spin susceptibility is written as the sum of two terms, $\chi \left( T,x \right)$ = $\chi_{\rm 1}(T,x) + \chi_{\rm 2}(T,x)$, where the first term ($\chi_{\rm 1}$) carries the pseudogap feature discovered by NMR shifts measurements, i.e., it is $T$-dependent far above $T_{\rm c}$. \cite{Alloul1989} The second susceptibility ($\chi_{\rm 2}$) is $T$ independent above $T_{\rm c}$, but it quickly disappears below $T_{\rm c}$, similar to what was concluded previously for La$_{1.85}$Sr$_{0.15}$CuO$_4$, and that reminds one of Fermi-liquid-like behavior.\cite{Haase2009} Another important experimental finding is the simple scaling behavior of the pseudogap susceptibility with the hole concentration in the underdoped regime (the pseudogap susceptibility changes proportionally with doping) that some of us anticipated earlier from linewidths studies.\cite{Rybicki2009} 

As a local, bulk probe NMR is not only powerful when it comes to measuring the uniform spin susceptibility. It is also very sensitive to inhomogeneities in the bulk of a material that manifest themselves in excessive broadenings of the NMR lines. However, since the HTSC are rather anisotropic, the lack of single crystals of sufficient size has hampered the investigation of the NMR linewidths. In addition, since most HTSC are non-stoichiometric, concomitant chemical disorder may induce linewidths that are of no great interest\cite{Eisaki2004}, while studies of broad lines are very time consuming. Nevertheless, there have been early accounts of unusual spatial inhomogeneities from NMR\cite{Haase2000} or nuclear quadrupole resonance (NQR)\cite{Singer2002}, but since stoichiometric materials like YBa$_2$Cu$_4$O$_{8}$ can have rather narrow lines\cite{Bankay1994} the importance of NMR linewidths studies of HTSC remained controversial, despite the fact that many other experimental methods indicated the presence of significant inhomogeneities.\cite{Jorgensen1988,Tranquada1995,Bianconi1996,Haskel1997,Yamada1998,Pan2001}

Here we will show that NMR linewidths in two studied single crystals require a two-component description, as well. We introduce a simple model based on a coherent variation of spin shifts and quadrupole splittings, which explains most of the NMR linewidths above about 60 K (for $T <$ 60 K we believe the linewidths are dominated by broadening due to the fluxoids). We find that the magnetic linewidth is dominated by a spatial variation of $\chi_{1}$, the susceptibility that carries the pseudogap. A smaller contribution from the variation of $\chi_{2}$ is also present, but abruptly disappears at higher $T$ (similar to what one finds in the case of motional narrowing). We also identify another $T$ independent broadening that appears for Cu as well as for Hg and is thus believed to be given by variations of spin, as well. Interestingly, this component appears to be coherent with the other broadening mechanisms for the optimally doped sample while being incoherent for the underdoped material.

Concurrent with the investigation summarized here, some of us developed a new kind of high-pressure NMR\cite{Haase2009a} and investigated the $^{17}$O NMR in YBa$_2$Cu$_4$O$_8$, which was shown to have single spin component behavior, earlier.\cite{Bankay1994} It was found that increasing pressure changes the uniform spin susceptibility in such a way that a two-component description is necessary for this material, as well.\cite{Meissner2011} The results also show that the second component ($\chi_{\rm 2}$) is rather weak at ambient pressure (that is why it was not identified with earlier measurements), but increases by an order of magnitude under high pressures where the NMR pseudogap has almost disappeared. 

It has been suggested that our two spin components are due to a (nodal) Fermi-liquid-like component and a (anti-nodal) spin liquid\cite{Barzykin2009}. Indeed, the presence of a Fermi-liquid-like component in the underdoped cuprates, which we have found with NMR is supported by other recent experiments. New dc resistivity measurements on HgBa$_{2}$CuO$_{4+\delta}$ and a careful analysis of experiments in a number of other cuprates showed that there is a doping and $T$-dependent region where the resistivity $\rho \propto T^2$.\cite{Barisic2012} There are two characteristic temperatures, $T^*$ below which resistivity deviates from linear dependence and $T^{**}$ below which $\rho \propto T^2$. The 'pseudogap' temperature $T^*$ has been recently associated with the novel magnetic excitation observed with neutron scattering experiments.\cite{Li2008a, Li2010, Li2011a, Li2012}
The temperature $T^{**}$ coincides with the maximum of the thermoelectric power.\cite{Yamamoto2000} Also optical spectroscopy measurements show that the near-nodal excitations	of the underdoped	cuprates obey a Fermi-liquid-like behavior.\cite{Mirzaei2012}  
Based on the NMR shift scaling we have found\cite{Meissner2011, Haase2012} it was suggested that electronic entropy data of YBa$_2$Cu$_3$O$_{6+x}$ and Bi$_2$Sr$_2$CaCu$_2$O$_{8+\delta}$ can be explained in a two-component scenario, as well\cite{Storey2012}. 

\section{Experimental}
\begin{figure}
\includegraphics[scale=0.5]{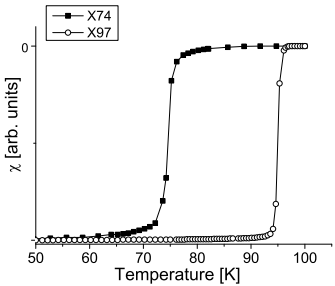}
\caption{\label{Fig:susceptibility} Temperature dependence of the uniform magnetic susceptibility, $\chi(T)$, for the underdoped (X74) and the optimally doped (X97) HgBa$_{2}$CuO$_{4+\delta}$ single crystals measured with zero field cooling.}
\end{figure}

The two crystals used in this study were annealed to result in one optimally doped crystal with $T_{\rm c}$ = 97 K (mass 30.3 mg) that we label X97 and one underdoped crystal with $T_{\rm c} $ = 74 K (mass 3.3 mg) that we label X74.\cite{zhao2006, Barisic2008} The uniform susceptibility data in Fig.~\ref{Fig:susceptibility} reveal very sharp phase transitions indicating high sample quality. 

$^{63}$Cu ($I$=3/2) and $^{199}$Hg ($I$=1/2) NMR measurements were typically carried out in a magnetic field $B_0$=11.75 T. For $^{63}$Cu the temperature ranged from 390 K to 20 K and for $^{199}$Hg from 295 K to 20 K. At low $T$ the signal loss due to small radio frequency (RF) penetration of the single crystal becomes excessive in connection with the decrease in the nuclear spin-lattice relaxation rate. However, above $T_{\rm c}$ we did not notice any unaccounted change in signal intensity. In order to study the origin of the $^{63}$Cu linewidth we also measured $^{63}$Cu NMR spectra between $B_0$=7.07 T and $B_0$=17.62 T. For $^{63}$Cu NMR most measurements were carried out using a frequency stepped spin echo technique for recording the broad lines with the pulse sequence $\tau_{\pi/2}-\tau-\tau_{\pi}$, where $\tau_{\pi/2}$ and $\tau_{\pi}$ are durations of the $\pi/2$ and $\pi$ pulse for the particular transition, respectively. The spin echoes were then integrated and plotted versus frequency. The Fourier transform (FT) of the spin echo was also used, but only when the line was narrow in comparison with the excitation/detection bandwidth. The $^{199}$Hg NMR line shapes were obtained from the FT of the second half of the spin echo excited with short RF pulses (a few $\mu$s). The pulse separation $\tau$ was 10-15 $\mu$s and 40 $\mu$s for $^{63}$Cu and $^{199}$Hg measurements, respectively (except for spin-echo decay measurements).

A particular nuclear transition was typically measured as a function of temperature ($T$) for a given orientation of the tetragonal single crystals with respect to the external magnetic field, in most cases with the field parallel ($c\parallel B_0$) or perpendicular ($c \bot B_0$) to the crystal $c$-axis. The mean resonance frequency of such a transition is denoted by $\nu_0$ and its mean magnetic shift by $K=(\nu_0-\nu_{\rm ref})/\nu_{\rm ref}$, where $\nu_{\rm ref}$ denotes the resonance frequency of a suitable reference compound: for $^{63}$Cu ($^{63}K$) we used fine metallic copper powder that has a Knight shift of 3820 ppm.\cite{Lutz1978} $^{199}$Hg shifts ($^{199}K$) were referenced to (CH$_3$)$_2$Hg using the referencing procedure described by Harris.\cite{Harris2008}

With more details of basic NMR theory for the cuprates available elsewhere,\cite{Slichter2007} here we repeat some fundamental definitions only. Both Cu isotopes, $^{63}$Cu and $^{65}$Cu, have nuclei with spin $I$=3/2, similar abundance, magnetic and quadrupole moments, and we focus here on $^{63}$Cu NMR. Its electric quadrupole moment eQ leads in the presence of an electric field gradient (EFG) $V_{ij}$ with the principle axis components $V_{\rm XX}, V_{\rm YY}, V_{\rm ZZ}$ and the asymmetry parameter $\eta  = \left( V_{\rm XX} - V_{\rm YY} \right) / V_{\rm ZZ},\,\,\,\left| {{V_{\rm ZZ}}} \right| \ge \left| {{V_{\rm YY}}} \right| \ge \left| {{V_{\rm XX}}} \right|$, to an electric quadrupole interaction. In the presence of a strong magnetic field $B_{0}$ the spin Hamiltonian in leading order can be written as \cite{Haase2004},
\begin{equation}
H = \hbar {\omega _0}{I_z} + {{3I_z^2 - I(I + 1)} \over {2I(2I - 1)}}eQ \cdot {1 \over 2} {{V_{zz}(\theta, \phi)}},
\label{eq:h1}
\end{equation}
with $\omega_{\rm 0}/2\pi=\nu_{\rm 0}$. $V_{zz}$ depends on the relative orientation $\theta, \phi$ of the nuclear quantization axis with respect to the symmetric tensor's principle axes system. We have with the quadrupole frequency $2\pi\nu_{\rm Q} ={\omega _{\rm Q}} \equiv {{3eQ{V_{ZZ}}} / {2I\left( {2I - 1} \right)}} \equiv {{3{e^2}qQ} / {2I\left( {2I - 1} \right)}}$,
\begin{equation}
\label{eq:h2}
H = \hbar {\omega _0}{I_z} + {{\hbar \omega _Q} \over 6}A_{\theta \phi}\left[ {3I_z^2 - I(I + 1)} \right],
\end{equation}
with
\begin{equation}
\label{eq:angle}
A_{\theta \phi}= {{{3{{\cos }^2}\theta  - 1} \over 2} + {\eta  \over 2}{{\sin }^2}\theta \cos 2\phi }.
\end{equation}
From the 2$I$+1 eigenvalues of (\ref{eq:h2}) we can write for the three Cu resonance lines (with m=-1/2 for the central transition), 
\begin{equation}
\label{eq:frequencies}
\nu_m (\theta,\phi) = {\nu_0} + \left( {m + {1 \over 2}} \right)\nu_{\rm Q}\cdot A_{\theta \phi}. 
\end{equation}
Since the quadrupole interaction for Cu in the cuprates can be rather large, higher order effects can be important and we employ a numerical diagonalization for fitting procedures. However, for most of this work second order corrections are sufficient for the shift of the central transition. 

For this tetragonal crystal and a partially filled Cu {3d$(x^2-y^2)$} orbital $V_{\rm ZZ}$ must be along the crystal $c$-axis, and we expect $\eta\approx0$ (this is indeed what we found experimentally for the X97 crystal earlier \cite{Rybicki2009}). Therefore, the two important orientations are $\theta = 0$ ($c\parallel B_{0}$) and $\theta = 90^0$ ($c\perp B_{0}$), i.e. the crystal $c$-axis is along the external field or perpendicular to it, respectively.

Similar arguments apply also to the total magnetic shift tensor, and we expect with $\eta=0$,
\begin{equation}
\label{eq:angular}
K \left( \theta  \right) = K_{\rm iso}  + K_{\rm ani} \left(3\rm{cos^2}\theta -1\right)/2,
\end{equation}
where $K_{\rm iso}$ and $K_{\rm ani}$ are the isotropic and anisotropic shift components, respectively. 

We have for the central transition (0) resonance frequencies up to 2nd order,
\begin{gather} 
\label{eq:central1}
\nu_{\parallel,0} = \nu_{\rm ref} \left( 1+ K_{\parallel} \right)\\
\label{eq:central1b}
\nu_{\perp,0} \approx \nu_{\rm ref} \left( 1+ K_{\perp} \right) +\frac{3}{16}\frac{\nu_{\rm Q}^2}{\nu_{\rm ref}}.
\end{gather} 
The two ($\pm$) satellite transitions for $c\parallel B_{0}$ are given by,
\begin{gather} \label{eq:satel1}
\nu_{\parallel,\pm}= \nu_{\rm ref} \left( 1+ K_{\parallel} \right) \pm \nu_{\rm Q},
\end{gather} 
and $K_{\parallel}=K_{\rm iso} +K_{\rm ani}$ and $K_{\perp}=K_{\rm iso}-K_{\rm ani}/2$ are the components of the magnetic shift tensor for the two orientations of the field. Since  $\nu_{\rm Q}$ can easily be determined, its contribution to the perpendicular shift can be removed.

For the $I$=1/2 $^{199}$Hg nucleus there is only the Zeeman term in the shift equations from above, however, since we find two Hg resonances they must have different shift tensors.

The total magnetic shifts for nucleus $n$ are then given by the following three terms,
\begin{equation}
{^nK} = {^nK_{\rm L}} + {^nK_{\rm S}}(T) + K_{\rm D} \left( T< T_{\rm c} \right),
\label{eq:shift}
\end{equation}
with the anisotropic orbital shift ${^nK}_{\rm L}$, the anisotropic spin (Knight) shift ${^nK}_{\rm S}$. $K_{\rm D}$ denotes the macroscopic diamagnetic shift (Meissner diamagnetism) that appears only below $T_{\rm c}$ in the mixed state; at our field strengths it is expected to be small on the $^{63}$Cu shift scale (we will estimate it below); note that $K_{\rm D}$ does not carry the label $n$ as it is independent of the gyromagnetic ratio for relative shift units. The orbital term $K_{\rm L}$ is $T$ independent while $K_{\rm S}$ can change significantly with $T$ in the cuprates (pseudogap).\cite{Alloul1989} 

\begin{figure}
\includegraphics[scale=0.75]{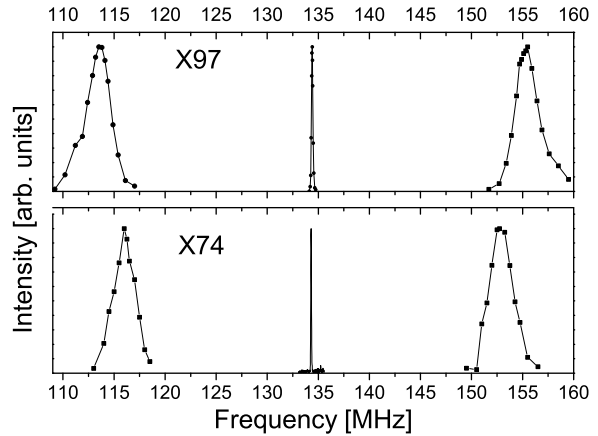}
\caption{\label{Fig:spectra}$^{63}$Cu NMR spectra at 11.75 T and at 100 K for $c\parallel B_{0}$ orientation for both single crystals showing central and satellite transition lines normalized to the same height in order to see the satellites better.}
\end{figure}

\section{Results}
\subsection{$^{63}$Cu NMR} \label{ResCu}
\begin{figure}
\includegraphics[scale=0.75]{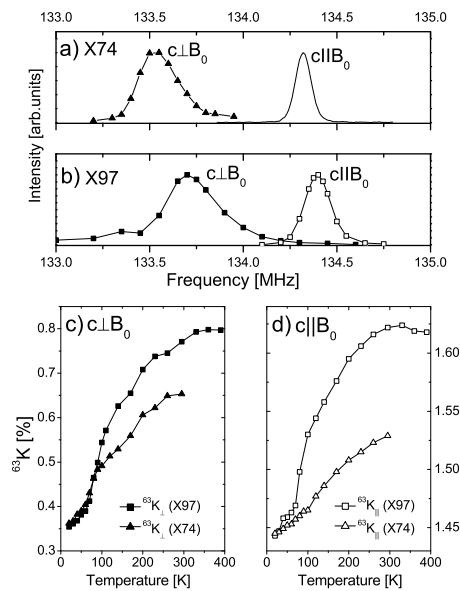}
\caption{$^{63}$Cu NMR central transition lines at 11.75 T for ${c\parallel B_{0}}$ and ${c\perp B_{0}}$ for a) X74 (at 120 K) and b) X97 (at 100 K) crystal, respectively. The ${c\parallel B_{0}}$ line shape for the X74 crystal shows no symbols since it is a FT of an echo signal while the other line shapes shown in a) and b) were obtained with the frequency stepped spin echo technique (see the Experimental for more details). Temperature dependence of $^{63}K$ for both crystals for c) $c\perp B_{0}$ (corrected for the second order quadrupolar contribution) and d) $c\parallel B_{0}$. Lines are to guide the eye.} 
\label{fig:Central}
\end{figure}

All three $^{63}$Cu NMR transitions, two satellite lines and one central line, for both single crystals for $c\parallel B_{0}$ and at 100 K are shown in Fig.~\ref{Fig:spectra}. All lines are normalized to the same height for a better display of the broad satellites (we find the expected intensity ratios conserved). For X97 (X74) the central transition has a center frequency of 134.41 MHz (134.29 MHz) and a linewidth of about 150 kHz (100 kHz); higher resolution plots of the central lines for $c\parallel B_{0}$ and $c\perp B_{0}$ and both crystals are shown in Fig.~\ref{fig:Central} (a,b).
For $\nu_{\rm Q}$ in (\ref{eq:satel1}) we find 20.88 MHz (18.46 MHz) for X97 (X74). The rather large satellite linewidth of 2.61 MHz (2.70 MHz) for X97 (X74) is in stark contrast to that of the central transition (about a factor of 20 smaller), but expected if the dominant mechanism is a distribution of EFGs that does not affect the central transition to first order.

The dependence of  $\nu_{\rm Q}$ and its distribution on $T$ is found to be rather weak (the change of lattice constants with $T$ is rather small\cite{Wagner1993}). For the X97 crystal we estimate from our data that $\nu_{\rm Q}$ changes by less than 1.4\% between 300 K and 50 K. For X74 we did not observe any $T$ dependent changes of $\nu_{\rm Q}$. 

The increase of $\nu_{\rm Q}$ with increasing hole concentration $x$ has been observed for HgBa$_{2}$CuO$_{4+\delta}$ before.\cite{Gippius1997} In fact, a linear increase in $\nu_{\rm Q}$ with $x$ has been found for all HTSC and is quantitatively accounted for by the increase in hole content, predominantly of the O $2{\rm p_\sigma}$ and Cu ${\rm 3d(x^2-y^2)}$ orbitals.\cite{Haase2004} For the X97 crystal we estimated from angular dependent measurements that the asymmetry parameter $\eta < 0.006$, in accord with the tetragonal crystal structure.\cite{Rybicki2009} We conclude that the satellite linewidths are given by largely symmetry conserving EFG variations.\\

\begin{figure}
\includegraphics[scale=0.75]{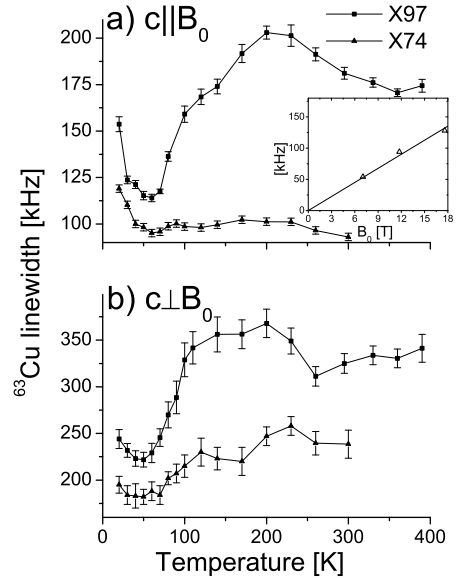}
\caption{\label{Fig:LWT}Temperature dependence of the $^{63}$Cu  central transition linewidths at 11.75 T for X97 and X74 crystals for a) ${c\parallel B_0}$ and  b) ${c\perp B_0}$ (lines are guides to the eye). Inset in a): field dependence of the linewidth for X74 measured for ${c\parallel B_0}$ at $T$=300 K.} 
\end{figure}

In order to investigate the origin of the linewidths of the central transitions we carried out measurements at various magnetic field strengths ($B_0$ = 7.05 T, 11.75 T and 17.62 T).  The results for the X74 crystal for ${c\parallel B_0}$ are shown in the inset of Fig.~\ref{Fig:LWT} (a). The linewidth in units of frequency is proportional to the field, as expected for magnetic broadening. The results for the X97 crystal are very similar (data not shown). The linewidth for $c\perp B_0$ is bigger than for ${c\parallel B_0}$ due to the distribution of the just mentioned quadrupole coupling, and we find experimentally at the highest $B_0$ and 300 K roughly a 100 kHz larger width compared to that for {$c\parallel B_0$}. 

The magnetic shifts and linewidths are strongly $T$-dependent. Fig.~\ref{fig:Central} (c,d) shows the $T$ dependence of $^{63}K$ for both crystals and for both orientations. These are the total magnetic shifts (only the second order quadrupolar contributions have been subtracted for $c\bot B_0$). We note that all $^{63}K$ shifts begin to decrease already far above $T_{\rm c}$ signaling pseudogap behavior. \cite{Alloul1989} For both single crystals there is no dramatic change in the shifts near $T_{\rm c}$. Our shift data are in agreement with what has been observed for this class of materials earlier and also from $^{17}$O NMR.\cite{Bobroff1997} 

In Fig.~\ref{fig:Central} (c,d) we observe for both orientations at the lowest $T$ that the $^{63}$Cu shifts are practically identical for both crystals. This is expected if the shifts at these $T$ are dominated by $^{63}K_{\rm L,\parallel}$ and $^{63}K_{\rm L,\perp}$, since the orbital shift has been found to be doping independent.\cite{Itoh1998,Ohsugi1994} With this assumption we find for the ratio $^{63}K_{\rm L,\parallel}/^{63}K_{\rm L,\perp}$=4.23, which is typical for HTSC.\cite{Renold2003} 

The $T$ dependences of the $^{63}$Cu central transition linewidths (full width at half height) for ${c\parallel B_0}$ and both crystals are presented in Fig.~\ref{Fig:LWT}a. The dependences for $c\perp B_0$ (Fig.~\ref{Fig:LWT}b) show qualitatively the same behavior as for ${c\parallel B_0}$, but the widths are bigger due to quadrupolar contribution. For the X97 crystal, starting from 400 K, the linewidths increase and reach a maximum at around 200 K, then they start to decrease. This trend continues also in the superconducting state down to about 60 K with perhaps a small jump at $T_{\rm c}$. Further decrease of $T$ results in an increase of the linewidth. For the X74 crystal the $T$ dependence of the linewidths is qualitatively similar. However, the relative change of the linewidth with $T$ is much smaller.

\subsection{$^{199}$Hg NMR}
\begin{figure}
\center{}
\includegraphics[scale=0.75]{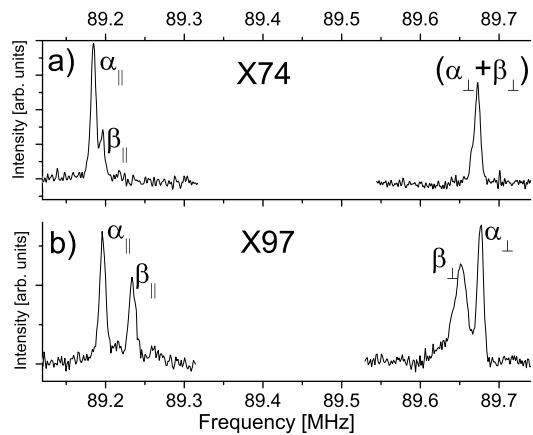}
\caption{$^{199}$Hg NMR spectra obtained at 295 K for the two single crystals a) X74 and b) X97 for $c\parallel B_0$ and $c \perp B_0$ orientation. The labels $\mathrm{\alpha}$ and $\mathrm{\beta}$ denote the two different Hg sites.}
\label{fig:spectra}
\end{figure}
\begin{figure}
\center{}
\includegraphics[scale=0.75]{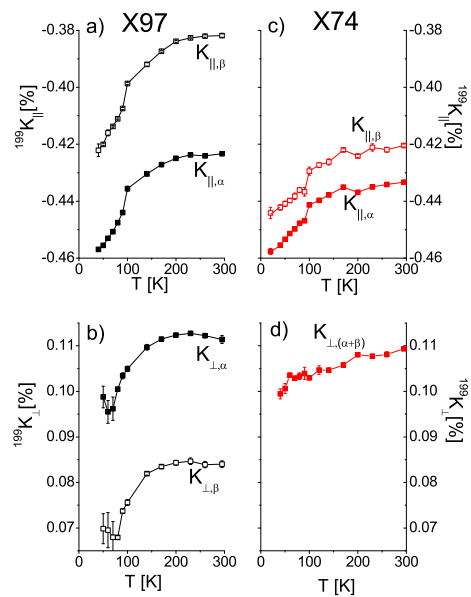}
\caption{Temperature dependence of the total $^{199}$Hg shifts, $^{199}$$K_{\parallel , \bot ,\alpha ,\beta } \left( T \right)$, for the X97 (a, b) and X74 crystal (c, d) with the magnetic field parallel to the $c$-axis, $c\parallel B_0$,(a, c), and perpendicular to the $c$-axis, $c \bot B_0$, (b, d). For the X74 crystal, the two resonances could not be resolved for $c\bot B_0$. (The full lines are guides to the eye.)} 
\label{fig:ShiftT}
\end{figure}

Typical $^{199}$Hg NMR spectra for both single crystals are shown in the upper and lower panel of Fig.~\ref{fig:spectra}.
$^{199}$Hg is a spin $I=1/2$ nucleus, i.e. no quadrupole interaction perturbs the Zeeman term (only a single resonance line is expected). However, we recognize in Fig.~\ref{fig:spectra}, in particular for the X97 crystal (lower panel), \textit{two} well-resolved Hg lines that we denote by $\alpha$ and $\beta$. These two lines must be due to non-equivalent Hg sites in the crystal lattice. The site that we call $\alpha$ is similar in both samples, not only in terms of intensity, but also the shifts at low $T$ are identical (see Fig.~\ref{fig:ShiftT}). We conclude that the $\beta$ line must be caused by doping with extra oxygens. Note that even for the X97 crystal both resonances are sharp, pointing to a well defined local structure for both sites. In previous studies on aligned powders, only a single broad asymmetric $^{199}$Hg line was reported.\cite{Suh1996a}

\begin{figure}
\center{}
\includegraphics[scale=0.75]{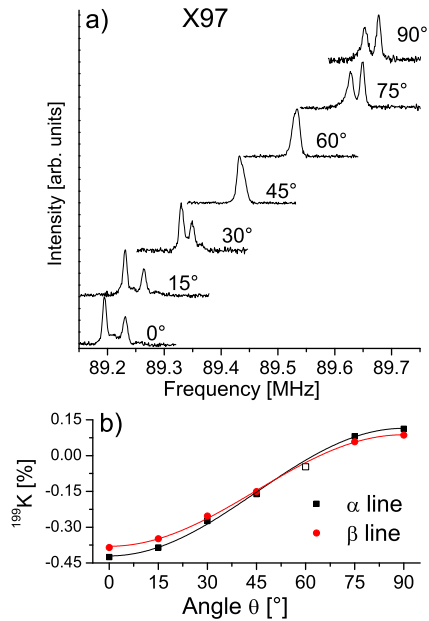}
\caption{Angular dependence of the $^{199}$Hg NMR spectra at 295 K for the optimally doped sample. a) Spectra at different angle $\theta$ between the magnetic field direction and the $c$-axis. b) Shifts $^{199}K$ vs. $\theta$ with fits (solid lines) to the Eq. \ref{eq:angular}. For $\theta$=60$^{\circ}$ we were unable to resolve two lines.}
\label{fig:AngularDependence}
\end{figure}

Since the doping may change the principal components of the shift tensor and/or its orientation, we recorded $^{199}$Hg spectra as a function of the angle $\theta$ between $B_0$ and the crystal $c$-axis. The data taken at 295 K for the X97 crystal are shown in Fig.~\ref{fig:AngularDependence}. Both Hg lines move within error according to a symmetric shift tensor whose main axis is the crystal $c$-axis. The $\beta$ Hg atoms whose orbital shift changes significantly with doping appear to retain this bonding symmetry. The anisotropy of the total magnetic shift in Fig.~\ref{fig:AngularDependence} is dominated by the orbital shift since its anisotropy is much larger than that of the spin shift, cf. also Fig.~\ref{fig:ShiftT}.

Expressing $^{199}K$ by Eq.~(\ref{eq:angular}) we find for the two sites $^{199}K_{\rm iso,\alpha }=\left( { - 0.064 \pm 0.004} \right)\%$, $^{199}K_{\rm ani,\alpha}=\left( { - 0.357 \pm 0.005} \right)\%$, $^{199}K_{\rm iso,\beta }  = \left( { - 0.068 \pm 0.003} \right)\%$, $^{199}K_{\rm ani,\beta}   = \left( { - 0.312 \pm 0.004} \right)\%$. The isotropic contribution remains the same while the anisotropy changes slightly with doping.


The $T$ dependences of the $^{199}$Hg shifts for both resonances in both crystals and two orientations are shown in Fig.~\ref{fig:ShiftT}. We see that they are very similar for both sites for a given crystal and orientation, but very anisotropic. Similar to $^{63}$Cu NMR and to other cuprates, the size of the $T$ dependent part is bigger for the sample with higher doping. From the $T$ dependence of $^{199}K(T)$ we can also conclude that the diamagnetic Meissner term ($K_{\rm D}$) that is expected below $T_{\rm c}$  must be small on the Cu shift scale: the total shift change, $^{199}K$, below $T_{\rm c}$ is less than $\approx$0.02\% and $\approx$0.01\% for $c\parallel B_0$ and $c\perp B_0$, respectively. Since $K_{\rm D}$ (in \%) is the same for all nuclei in the sample, we see that these values are very small on the $^{63}K$ shift scale. Due to the poor signal-to-noise ratio in the superconducting state for these small single crystals (shielding effects and exceedingly slow nuclear relaxation) we could not reliably measure signals at lower temperatures. The upturn for $^{199}K_{\perp}$  at low temperatures is due to a broad component that is not found in powders \cite{Suh1996a} and must be due to the increasing influence of surface effects as the temperature decreases. 

\begin{figure}
\center{}
\includegraphics[scale=0.6]{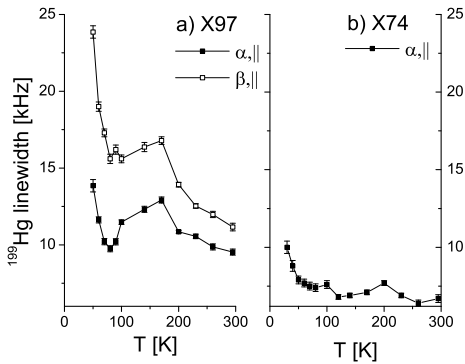}
\caption{Temperature dependence of $^{199}$Hg linewidth for \mbox{$c\parallel B_0$} orientation for a) the X97 crystal $\alpha$ and $\beta$ line, b) the X74 crystal $\alpha$ line. The linewidth for $\beta$ line could not be calculated reliably (cf. Fig. \ref{fig:spectra}a).}
\label{fig:199KLW}
\end{figure}

The $^{199}$Hg NMR linewidths for $c\parallel B_0$ for X97 and X74 are shown in Fig.~\ref{fig:199KLW}. We see that they are much narrower compared to Cu $c\parallel B_0$, for which the broadening is also only of magnetic origin. The $T$ dependences are similar to that for the central transition of Cu. This is expected if the changes of the linewidths with $T$ are due to inhomogeneities of the field induced electronic spin polarization. For $c\perp B_0$ and X97 (data not shown) the linewidth of the $\alpha$ line shows a $T$ dependence similar to $c\parallel B_0$ while for the $\beta$ site it increases as temperature decreases. Due to line overlap we were unable to reliably measure the linewidths for $c\perp B_0$ for the X74 crystal. 

At room temperature for  X97 the spin-spin relaxation was approximately Gaussian with $1/T_{G}\approx$ 1.8 ms$^{-1}$, which is much bigger than expected from the direct nuclear dipolar interaction that amounts to about \mbox{0.2 ms$^{-1}$}.\cite{Suh1996a} Thus, the observed linewidths are all inhomogeneous.

\section{Discussion}
\subsection{The two $^{199}$Hg resonance lines}

The presence of a second, sharp Hg line despite the large Hg orbital shift points to a very well defined crystallographic site. At low $T$, where we can assume similar diamagnetic terms for both crystals, the orbital shift of the $\alpha$ site, $^{199}K_{\rm L,\alpha,\parallel}$, is independent of doping while $^{199}K_{\rm L,\beta,\parallel}$ increases with doping and indicates a well defined change of all Hg atoms affected by the doping. For X97 we find ${^{199}K_{\rm L,\beta,\parallel}}-{^{199}K_{\rm L,\alpha,\parallel}}\approx 0.035\%$ and for X74, ${^{199}K_{\rm L,\beta,\parallel}}-{^{199}K_{\rm L,\alpha,\parallel}}\approx 0.015\%$. That is, the orbital shift for $c\parallel{B_0}$ grows by 0.02\% (only about 4\% of the total orbital shift).
For X97 and $c\bot{B_0}$ we have a similar change of the orbital term, $^{199}K_{\rm L,\bot,\beta}-^{199}K_{\rm L,\bot,\alpha}\approx 0.025\%$ , for X74 and $c\bot{B_0}$ the two sites could not be resolved.

The shift anisotropy of the Hg $\alpha$ site is expected for Hg atoms with axial symmetry. If the doped, extra oxygen enters the Hg plane in the middle of a Hg plaquette,\cite{Wagner1993, Huang1995} it may induce a well-defined Hg $\beta$ site in accordance with what we find. The symmetry of the $\beta$ site shift tensor is similar to that of the $\alpha$ site so the extra oxygen does not change the wave functions, and the changes of the principal components are probably due to changes in energy.

From the relative line intensities we can roughly estimate the extra oxygen content $\delta$ for our samples assuming that the extra oxygen is O$_{\delta} ^{2+}$. From the spectra with ${c \parallel B_0}$ at higher temperatures where the lines are narrow we estimate the relative line intensities of $I_{\alpha} / I_{\beta} \approx$ 3.3$\pm$0.3 and 1.4$\pm$0.2 for X74 and X97, respectively. Assuming a stoichiometric composition with O$_{\delta}$ located in the center of the Hg plaquette, these values of $I_{\alpha} / I_{\beta}$ correspond to $\delta=0.07\pm0.01$ (X74) and $\delta=0.15\pm0.01$ (X97). These values are in good agreement with a formula suggested by Balagurov et al.,\cite{Balagurov1999} \mbox{$T_{\rm c}=T_{\rm c,max} \left[ 1-52 \left( \delta - 0.127 \right)^2 \right]$}. For the values of $T_{\rm c}$ of our crystals we obtain $\delta=0.06$ and $\delta=0.13$. 

It has been found that Hg vacancies can be present in HgBa$_{2}$CuO$_{4+\delta}$.\cite{Antipov2002, zhao2006} Such a defect also affects 4 Hg atoms and could in principle produce a distinct NMR line due to Hg atoms that are close to a vacancy. If Hg vacancies are present $\delta$ will be lower than calculated above from the measured line intensities. With an increasing amount of Hg vacancies the theoretical ratio $I_{\alpha} / I_{\beta}$ decreases. The values of $\delta$ calculated above for a stoichiometric material are indeed slightly higher than expected from the formula given by Balagurov et al. After close inspection of the $^{199}$Hg spectrum for the X97 crystal (Fig.~\ref{fig:spectra}b) one can notice weak signals (i.e. a broad foot) for $c \perp B_0$, which could be due to Hg vacancies, Hg atoms with two O$_{\delta}$ as neighbors or other types of defects. 

It has been suggested that O$_{\delta}$ changes position in the lattice as the extra oxygen content increases.\cite{Jorgensen1997} However, precise evaluation of the nature of ${^{199}K_{\rm L,\beta}}$ change with $\delta$ would require studying more samples with different extra oxygen content.

\subsection{$^{63}$Cu Quadrupole Splittings}

The largely $T$ independent quadrupole interaction for Cu with ${\eta \approx 0}$ is typical for HTSC, and the values for the crystals investigated here are in agreement with what has been reported before. \cite{Gippius1997, Rybicki2009} It reflects the fact that Cu is near a $3\rm d^9$ configuration with a hole in the ${\rm 3d(x^2-y^2)}$ orbital that is hybridized with the surrounding four oxygen ${\rm 2p_\sigma}$ orbitals that predominantly take up the holes upon doping.\cite{Haase2004} This causes quadrupole splittings or electric field gradients (EFG) that depend approximately linearly on the extra oxygen content and thus on the hole concentration in the CuO$_2$ plane. While for planar O the EFG is given by the ${\rm 2p_\sigma}$ holes only, for Cu it depends on the Cu as well as O holes. With no $^{17}$O NMR splittings available, we cannot determine the number of holes that enter Cu ${\rm 3d(x^2-y^2)}$ upon doping, but the slope of $\nu_Q$ vs. $x$ of about 48 MHz/hole suggests that some fraction of holes also enter ${\rm 3d(x^2-y^2)}$.\cite{Haase2004} 

As mentioned earlier, the rather large linewidths of the $^{63}$Cu satellites shown in Fig.~\ref{Fig:spectra} are dominated by a distribution of quadrupole splittings, and thus a distribution of EFGs and they are similar to those observed by NQR (i.e. at zero magnetic field).\cite{Gippius1997} Given the superior chemical homogeneity of the crystals reflected in the very narrow $^{199}$Hg NMR lines, this may seem a surprising result. Indeed, if one tries to relate the linewidths to local structural disorder (buckling of the CuO$_2$ plane) static tilts of unphysical size (about 20 degrees) would be necessary and simple local disorder does not explain our findings. Likewise, the doped O$^{2+}$ ions in the Hg planes with their Coulomb potential do not produce sufficiently large gradients.\cite{Haase2004} In addition, such disorder likely comes with $\eta \sim 1$, but we find $\eta < 0.006$. However, since the local charge density has a strong effect on the EFG an inhomogeneity in this charge density (e.g., triggered by structural disorder or the dopant) can explain the observed linewidths. A variation of the Cu hole density of less than 0.03 would account for the observed linewidths. However, for various other non-stoichiometric HTSC a substantial variation of the charge in the oxygen ${\rm 2p_\sigma}$ orbitals has been found.\cite{Haase2000, Haase2001, Haase2002}  This inhomogeneity in the ${\rm 2p_\sigma}$ hole density causes changes of the EFG at the Cu site, that can explain the total Cu satellite linewidth.\cite{Haase2004} While we have no $^{17}$O data available for the crystals under study, given the size of the satellite widths, it appears likely that a charge density variation of holes in the ${\rm 2p_\sigma}$ orbitals dominates the Cu EFG variation for this material, as well. Recent Hartree-Fock calculations seem to confirm that the Coulomb potential of the dopant is responsible for the charge density variation.\cite{Chen2011}  

\begin{figure}
\center{}
\includegraphics[scale=0.75]{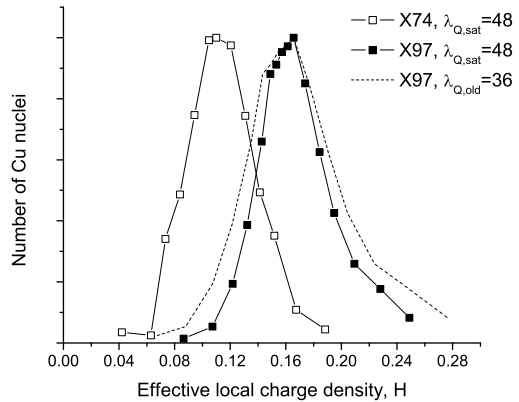}
\caption{The quadrupolar satellite line shapes plotted against the effective local charge density described by parameter $H$ for both single crystals, see text for more details.}
\label{fig:satellites}
\end{figure}

Since we cannot say with certainty how the holes are distributed between Cu and O, we define an effective local charge density $H$ (i.e., $\left<H\right>\equiv x$). Then, the linewidth of the $^{63}$Cu satellites would correspond to a spatial variation of $H$ in the CuO$_2$ plane. We write, ${\nu_{\rm Q}=\nu_{\rm Q,0}+{\rm \lambda_{Q,sat}}\cdot \left<H\right>}$, where ${\rm \lambda_{Q,sat}}$ is a doping independent constant that can be extracted from experiment.\cite{Haase2004} From an NQR study on powder samples\cite{Gippius1997} we estimated earlier ${\rm \lambda_{Q,old}}$  = 36 MHz per CuO$_2$ hole and used this value in a previous study.\cite{Rybicki2009} However, the phase diagram derived for single crystals\cite{Barisic2008} appears somewhat different on the underdoped side and from T$_{\rm c}$ values we estimate $\left< H\right>_{X74}$=0.11 and $\left< H\right>_{X97}$=0.16 for the X74 and X97 crystal, respectively. With these values  we obtain a larger value of ${\rm \lambda_{Q,sat}}$  = 48 MHz per CuO$_2$ hole, which will be used throughout this paper. We can then convert the satellite frequency to an effective local charge density described by parameter $H$. Since the NMR intensity is proportional to the number of nuclei, the satellite line shapes represent histograms of the local charge density in the CuO$_2$ plane. Such plots are shown in Fig.~\ref{fig:satellites} for both crystals for ${\rm \lambda_{Q,sat}}$=48 MHz/hole and ${\rm \lambda_{Q,old}}$=36 MHz/hole for the X97 crystal for comparison. From Fig.~\ref{fig:satellites} we infer rather strong variations of $H$, as found for other HTSC.\cite{Haase2000, Haase2002, Singer2002} Static variations of the charge density are typically much smaller only in stoichiometric compounds. For YBa$_2$Cu$_4$O$_8$ linewidths as low as 125 kHz (NQR on powder sample)\cite{Raffa1999, Mali2002} and 100 kHz (NMR on single crystal)\cite{Meissner2012}, for nearly stoichiometric YBa$_2$Cu$_3$O$_{7-\delta}$ linewidths of 200 kHz (NQR on powders)\cite{Schiefer1989, Klein1997} and 90 kHz (NMR on single crystal)\cite{Haase_u} were observed. However, for Tl$_{2}$Ba$_{2}$CuO$_{6+\delta}$, which is structurally very similar to HgBa$_{2}$CuO$_{4+\delta}$ NQR linewidths of about 2.5 MHz were measured.\cite{Kambe1991, Itoh2006} 
With ${\rm \lambda_{Q}}$=48 MHz/hole the linewidths (as full width at half maximum) in Fig.~\ref{fig:satellites} correspond to distributions of $H$ of about 0.058 for both crystals. Similar charge density variations have been observed, e.g., for La$_{2-x}$Sr$_x$CuO$_4$. \cite{Singer2002}
Note that we do not observe singularities in the histograms (cf. Fig.~\ref{fig:satellites}) as one might expect from highly ordered charge density variations, but one can think of various scenarios that will smooth such singularities.

\subsection{Magnetic Shifts}
While the total shifts are similar to those of other HTSC, the samples investigated here have sizable $T$-dependent  $^{63}$Cu shifts also for ${c\parallel{B_0}}$ ($^{63}K_\parallel(T)$), cf.~Fig.~\ref{fig:Central}. This is similar to Tl$_{2}$Ba$_{2}$CuO$_{y}$,\cite{Kambe1993} but very different from what one finds for YBa$_2$Cu$_3$O$_{7-\delta}$ and La$_{2-x}$Sr$_x$CuO$_4$, which only have T-dependent $^{63}K_\bot(T)$.\cite{Takigawa1989, Barret1990b, Takigawa1991, Ohsugi1994} In addition, optimally doped YBa$_2$Cu$_3$O$_{7-\delta}$ and La$_{2-x}$Sr$_x$CuO$_4$ show $^{63}K_\bot(T)$ that are rather flat in $T$ above $T_{\rm c}$ and drop rapidly below.\cite{Takigawa1989, Ohsugi1994} Quite to the contrary, $^{63}K_{\rm \bot, X97}(T)$ shows pseudogap behavior, i.e., a strong $T$ dependence above $T_{\rm c}$. The $^{17}$O NMR of optimally doped YBa$_2$Cu$_3$O$_{7-\delta}$ does not show pseudogap behavior for $^{17}K_\parallel(T)$, contrary to optimally doped La$_{2-x}$Sr$_x$CuO$_4$ where ${^{17}K_\parallel(T)}$ is similar to our $^{63}K(T)$.\cite{ Horvatic1989, Ishida1991}

In a separate account\cite{Haase2012} we discussed the magnetic shifts for X74 and X97. We found that they must be explained in a two-component scenario in which two spin components with different $T$ dependences couple with different hyperfine coupling coefficients to the same nucleus, as discovered recently in two other systems.\cite{Haase2009, Meissner2011} The shifts are then given by,
\begin{equation}
\label{eq:twocompshift1}
{^nK}_{\parallel , \perp}(x,T)={^np}_{\parallel , \perp}\cdot \chi_1(x,T)+{^nq}_{\parallel , \perp}\cdot \chi_2(x,T).
\end{equation}
It was concluded that each susceptibility consists of two terms, i.e., ${\chi_1=\chi_{\rm AA} +\chi_{\rm AB}}$ and  ${\chi_2=\chi_{\rm BB} +\chi_{\rm AB}}$, and it follows,
\begin{equation}
\label{eq:twocompshift2}
{^nK}_{\parallel , \perp}={^np}_{\parallel , \perp}\cdot (\chi_{\rm AA}+\chi_{\rm AB})+{^nq}_{\parallel , \perp}\cdot (\chi_{\rm AB}+\chi_{\rm BB}).
\end{equation}

Furthermore, it was shown that $\chi_{\rm AA}(x,T)$, which is responsible for the pseudogap feature in the NMR shifts, appears to obey scaling behavior (below 300 K): $\chi_{\rm AA}(x_j,T)=(x_j/x_i) \cdot \chi_{\rm AA}(x_i,T)$. Contrary to $\chi_{\rm AA}$, which does not vanish rapidly below $T_{\rm c}$, the other two susceptibilities $\chi_{\rm AB}$ and $\chi_{\rm BB}$ are $T$ independent above $T_{\rm c}$, but vanish rapidly at $T_{\rm c}$. This scenario follows from the Cu as well as Hg shifts. 
The corresponding hyperfine coefficients were found to be independent of doping and for HgBa$_{2}$CuO$_{4+\delta}$ we find: ${^{63}p}_\parallel /{ ^{63}p}_\bot \approx 0.4$, ${^{199}p}_\parallel /{^{63}p}_\parallel \approx 0.12$, \mbox{$(\chi_{\rm AB, X97}+\chi_{\rm BB, X97})\approx 0$}, $\chi_{\rm AA, X97} / \chi_{\rm AA, X74}\approx 1.4$ and $^{63}p_{\parallel} \cdot \Delta \chi_{\rm AB, X97} \approx 0.06\%$. \cite{Haase2012}  

From the shift data one can make further estimates. From Fig.~\ref{fig:Central} (c,d) we can read off the total change of the shifts $\Delta ^{63}K(T)$ with temperature. We find $\Delta ^{63}K_{\rm \parallel,X97}\approx 0.178\%$ and therefore \mbox{$^{63}p_{\parallel} \cdot \chi_{\rm AA, X97}(300K) \approx 0.12\%$}. This allows us to estimate $^{63}p_{\perp} \cdot \chi_{\rm AA, X97}(300K) \approx 0.3\%$, but from Fig.~\ref{fig:Central}(c) we know that $\Delta ^{63}K_{\rm \perp,X97}\approx0.418\%$. We conclude that \mbox{$^{63}p_{\perp} \cdot \chi_{\rm AB, X97} \approx 0.12\%$}. Using the scaling of $\chi_{\rm AA}$ with doping we can now make similar estimates for the X74 crystal. We find \mbox{$^{63} p_{\parallel} \cdot \chi_ {\rm AA, X74} \approx 0.086\%$}, which is equal to $\Delta ^{63}K_{\rm \parallel,X74}$ obtained from Fig.~\ref{fig:Central}(d). We also find \mbox{$^{63} p_{\perp} \cdot \chi_ {\rm AA, X74} \approx 0.21\%$}, from Fig.~\ref{fig:Central}(c) we have $\Delta ^{63}K_{\rm \perp,X74}=0.3\%$ and therefore we conclude that \mbox{$\left(^{63}p_{\perp} + ^{63}q_{\perp} \right) \chi_{\rm AB, X74} + ^{63}q_{\perp} \chi_{\rm BB, X74} \approx 0.09\%$}. 

In order to summarize and simplify the notation for the further discussion we define,
\begin{equation}
\label{eq:simple}
A\equiv \chi_{\rm AA}, \hspace{0.2cm} B\equiv \chi_{\rm BB}, \hspace{0.2cm} C\equiv \chi_{\rm AB}.
\end{equation}
We then have for the two crystals,
\begin{subequations}
\label{eq:set}
\begin {align}
K_{\parallel, X74}&\approx{p}_\parallel  A_{\rm X74 } \\
K_{\rm  \bot, X74}&\approx{ p}_\bot  A_{\rm X74}+\left({p}_\bot+{q}_\bot\right) C_{\rm X74}+ {q}_\bot B_{\rm X74}\\
K_{\rm  \parallel, X97} &\approx {p}_\parallel \cdot \left[A_{\rm X97} + C_{\rm X97}\right]\\
K_{\rm \bot, X97} &\approx {p}_\bot \cdot \left[A_{\rm X97} + C_{\rm X97}\right].
\end{align}
\end{subequations}
For $^{63}$Cu and above $T_{\rm c}$ with the constants estimated above it follows:
\begin{subequations}
\label{eq:set2}
\begin {align}
K_{\parallel,X74} & \approx{p}_\parallel  A_{\rm X74} \\ 
K_{\rm \bot, X74}& \approx {p}_\bot A_{\rm X74} +0.09\% \\ 
K_{\rm \parallel, X97} & \approx {p}_\parallel A_{\rm X97}+0.06\% \\
K_{\rm \bot, X97} & \approx {p}_\bot A_{\rm X97} + 0.12\%
\end{align}
\end{subequations}
with the constant terms vanishing quickly at $T_{\rm c}$.

\subsection{$^{63}$Cu and $^{199}$Hg Linewidths} 
The widths of the central $^{63}$Cu lines of both crystals are doping and temperature dependent, cf.~Fig.~\ref{Fig:LWT}. For comparison, in La$_{2-x}$Sr$_x$CuO$_4$ the $^{63}$Cu linewidths increase with decreasing temperature (similar to what was found for YBa$_2$Cu$_4$O$_8$, but here the widths are much smaller).\cite{Haase2000, Bankay1994} However, the linewidth of planar $^{17}$O in La$_{1,85}$Sr$_{0.15}$CuO$_4$ shows a more complicated $T$-dependence that is similar to that of  $^{63}$Cu in our X97 sample.\cite{Haase2000} 

As noted previously from measurements at different fields we confirmed that the linewidths for $^{63}$Cu and ${c\parallel{B_0}}$ are entirely magnetic (cf. chapter \ref{ResCu} and inset in Fig. \ref{Fig:LWT}a). 
For $c\bot{B_0}$ we have higher order quadrupolar contributions to the $^{63}$Cu widths. 

Since all linewidths are  inhomogeneous (typical room temperature spin echo decay constants are less than 100 $\mathrm{\mu s}$ for Cu and about 580 $\mathrm{\mu s}$ for Hg), they could simply be due to a spatial distribution of the field induced spin polarization. We already know that the effective local charge density ($H$) varies across the CuO$_2$ plane and with it the quadrupole splitting $\nu_{\rm Q}(H)$. Furthermore, we know that $A_X\equiv \chi_{\rm AA}$ is a function of the average hole concentration $x\equiv \left<H\right>$. 

Therefore, in order to investigate the linewidths more quantitatively we introduce a simple model that assumes coherent variations of all shift contributions as a function of $H$. We show later, from comparison of the ${^{63}}$Cu widths with those of ${^{199}}$Hg that we can neglect orbital shift variations as a substantial source of linewidth. We thus consider only variations of the spin shift and where appropriate a quadrupolar term $K_{\rm Q}$, and write,
\begin{equation}
\label{eq:ShiftSimpleH}
K(H)={p} A(H)+{(p+q)} C(H)+{q} B(H)+K_{\rm Q}(H).
\end{equation}
For $c \bot B_0$ the quadrupolar contribution (that depends on $B_0$) can be calculated from \eqref{eq:central1b}, or numerically. It will be approximately linear in $H$. In fact, we will use a linear approach for all components $Y(H)$ in \eqref{eq:ShiftSimpleH}, i.e., $Y(H)\approx Y(\left< H \right>)+y\cdot (H-\left< H \right>)$, where $y$ is the derivative of $Y$ with respect to $H$ near $\left< H \right>$. With $h\equiv H-\left< H \right>$ we write for the deviation of the shift from its mean value, $\delta K(h) \equiv K(H)-K(\left<H\right>)$,
\begin{equation}
\label{eq:ShiftDerivative}
\delta K(h)=\left[{p}\cdot a+{(p+q)}\cdot c+{q} \cdot b +\lambda_{\rm Q}\right]\cdot h.
\end{equation}
Then, the root mean square (RMS) shift variation that is responsible for the linewidth is given by,
\begin{equation}
\label{eq:Linewidth}
\sqrt{\left< \delta K^2 \right>}=\left|{p}\cdot a+{(p+q)}\cdot c+{q} \cdot b +{\rm \lambda_Q}\right|\cdot \sqrt{\left< h^2 \right>}.
\end{equation}

For the satellite transitions the linewidths are dominated by quadrupole interaction and the RMS widths are given by  $\sqrt{\left< \delta K_{sat}^2 \right>}=\left|\lambda_{\rm Q,sat}\right|\cdot \sqrt{\left< h^2 \right>}$. For $c\parallel B_0$ we determined $\lambda_{\rm Q,sat} \approx$ 48 MHz/hole, and we found from the linewidths in Fig.~\ref{fig:satellites} (assuming $\delta \nu=2\sqrt{\left< h^2 \right>} \sqrt{2 \ln(2)}$) practically the same RMS of $h$ for both crystals X74 and X97, 
\begin{equation}
\label{eq:h}
\sqrt{\left< h^2 \right>}~\approx0.025.
\end{equation}
Using $\lambda_{\rm Q,old}\approx$ 36 MHz/hole we get $\sqrt{\left< h^2 \right>}~\approx$0.033. 

Since $A_X$ in \eqref{eq:simple} is the only $T$-dependent term above $T_{\rm c}$, we only expect a $T$-dependent linewidth for ${T>T_{\rm c}}$ from $a(T)$ in \eqref{eq:Linewidth}. We found earlier that $A_{X97}(T) = 1.4 \cdot A_{X74}(T)$, i.e., both susceptibilities are proportional to each other with the constant given by the ratio of the average hole concentration.\cite{Haase2012} Therefore, we set $a_X(T)=A_X(h,T)/ \left< H \right>_{X}$, and we can express ${p}A_X$ by the experimentally measured shift. We conclude that the linewidths must then be linear functions of the experimentally measured shifts for {$T>T_{\rm c}$. In addition, with the rapid disappearance of $C_X$ and $B_X$ at $T_{\rm c}$ also their derivatives $c_X$ and $b_X$ are expected to vanish, and width and shift should be linear functions of each other below $T_{\rm c}$, as well. We thus write,
\begin{equation}
\label{eq:LinewidthEqn}
\begin{split}
\sqrt{\left< \delta K^2 \right>}=&|\lambda_{\rm Q} + K/\left<H\right>-\left[{(p+q)}C+{q}B\right]/\left<H\right>\\
&+({p+q})c+{q}b|\cdot \sqrt{\left<h^2 \right>}.
\end{split}
\end{equation}

\subsubsection{X97 Linewidths}
We now discuss the experimental results for the X97 crystal. From \eqref{eq:LinewidthEqn} we have with the shift from \eqref{eq:set},
\begin{equation}
\label{eq:LinewidthX9702}
\begin{split}
&\sqrt{\left< \delta K_{\rm X97}^2 \right>}=|\lambda_{\rm Q,X97} + K_{\rm X97}/\left<H\right>_{\rm X97}-{p}C_{\rm X97}/\left<H\right>_{\rm X97}\\
&+{p}c_{\rm X97}+{q}[c_{\rm X97}+b_{\rm X97}]|\cdot \sqrt{\left<h^2 \right>}.
\end{split}
\end{equation}

The corresponding experimental linewidth vs.\ shift plots are shown in Figs.~(\ref{fig:63LWK_X97}) and (\ref{fig:Hg_WvsK}) for $^{63}$Cu and $^{199}$Hg, respectively. We find indeed straight lines with fixed slopes in the central region. In addition, common to all four plots are also high-$T$ drop-offs, low-$T$ upturns, and also a $T$ independent offset that is not contained in our model.\\

\textbf{$^{63}$Cu}: For $^{63}$Cu and $c\parallel{B_0}$ in Fig.~\ref{fig:63LWK_X97} (a), we find a linear dependence between about 60 K and 200 K without a rapid change at $T_{\rm c}$, but we know from the shifts that $C_{\rm X97}$ does change at $T_{\rm c}$ (it can be seen in Fig.~\ref{Fig:LWT}). Consequently, the change in $C_{\rm X97}$ must be offset by a change in $c_{\rm X97}$ (and/or $b_{\rm X97}$). From the linear slope we determine $\sqrt{\left< h^2 \right>}/\left<H\right>_{\rm X97} \approx$ 0.22, in good agreement with what can be calculated from the satellite width (0.16 and 0.21 for $\lambda_{\rm Q,sat}$ and $\lambda_{\rm Q,old}$, respectively). This says that the variation of the linewidth with $T$ in this linear region is dominated by a substantial spatial variation of ${ A_X} \equiv \chi_{\rm AA}$.
STM, which is a surface probe, also indicates that the inhomogeneity comes mostly from a spatial variation of the pseudogap while the superconducting gap is more homogeneous.\cite{Hoffman2002, McElroy2005, Boyer2007, Pushp2009, Parker2010, Schmidt2011, Zeljkovic2012}

At high $T$ we observe a drop in the linewidth that causes a deviation from the linear behavior. This is rather unusual since the shift still increases, cf.\ Fig.~\ref{fig:Central}. With our model we conclude that $c_{\rm X97}$ and/or $b_{\rm X97}$ begin to vanish so that we see a drop due to [$-{\rm ^{63}p_\parallel}C_{\rm X97}\approx -0.06\%$]. We expect a change of [$-{\rm ^{63}p_\parallel}C_{\rm X97}\cdot 0.22 \approx -0.013\%$], or about 20~kHz at 11.75 T, in modest agreement with the linewidth that shows a decrease of $\approx$31~kHz. A vanishing of the derivatives means the inhomogeneities in $C_{\rm X97}$ and/or $B_{\rm X97}$ disappear while the average shift still grows.

The low $T$ increase we believe is caused by the freezing of the fluxoid lattice since the Hg data show a corresponding change as well. We will discuss the low $T$ region separately below. 

Unexplainable with our model \eqref{eq:LinewidthX9702} is a substantial low $T$ intercept $\sqrt{\left< \delta K_{\rm X97}^2 \right>}(T\rightarrow0)$ for zero shift, which we find to be about 100 kHz or 0.076\% while we expect it to be zero. Note that it must be a constant also in the linear region and probably beyond it. In addittion, if it was incoherent with the other broadening mechanisms in \eqref{eq:LinewidthX9702}, these would hardly be visible in a large incoherent background. We will call the associated linewidth $\zeta$. This constant broadening ($^{63}\zeta_{\parallel,\rm X97} \approx$ 100 kHz) does not appear to be of orbital origin since we show below that it is of similar size at the Hg nucleus. In separate field-dependent experiments we verified that this large background broadening is of magnetic origin. $\zeta$ is thus based on a modulation of the spin density with a wavevector different from zero as it does not contribute to the average shift.
\begin{figure}
\center{}
\includegraphics[scale=0.75]{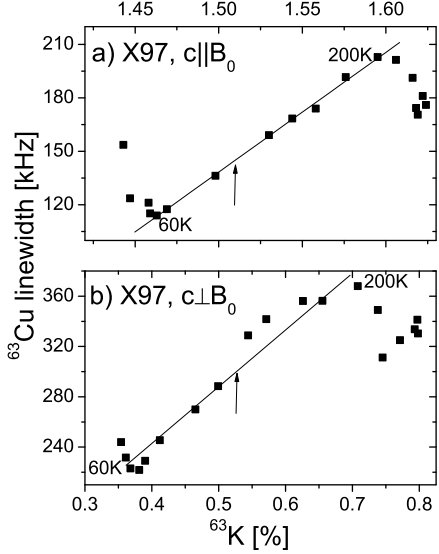}
\caption{$^{63}$Cu central transition linewidth for the X97 crystal as a function of the shift, $^{63}K$ with temperature as an implicit parameter for a) $c\parallel B_0$ and b) $c\perp B_0$. The straight lines are fits and the arrows indicate $T_{\rm c}$.}
\label{fig:63LWK_X97}
\end{figure}

For $c\bot{B_0}$ we observe a similar scenario with a similar constant slope ($\sqrt{\left< h^2 \right>}/\left<H\right>_{\rm X97} \approx$ 0.18) in the same $T$ range, except for a sudden drop of about 0.013\% in the offset at $T_{\rm c}$, expected in general from the change of the high-$T$ constants $C_X, B_X$. The high-$T$ drop is $\approx$50 kHz (0.038\%), and thus in agreement with the expected anisotropy of ${^{63}p_\parallel/^{63}p_\bot}\approx 0.4$. We calculate ${\rm \lambda_{Q,X97}} \approx$ +2.8 MHz/hole at 11.75~T, from which an intercept of $\sqrt{\left< \delta K_{\rm X97}^2 \right>}(T=0) \approx 165$ kHz (at 11.75 T) results, while we find about 210 kHz from experiment. This means the additional broadening is $^{63}\zeta_{\rm \bot,X97}\approx$45 kHz.

The $^{63}$Cu linewidths give a consistent picture of coherent broadening (above the fluxoid freezing) in which a spatial variation of the charge density $H$ is the variable that triggers a variation of the quadrupole interaction and the various shift contributions. In addition, there is a large $T$ independent broadening present that has an anisotropy of about 2.2, i.e., $^{63}\zeta_{\rm \parallel,X97}/^{63}\zeta_{\rm \bot,X97}\approx 2.2$. Note that wavelength effects with anisotropic hyperfine scenarios affect the anisotropy so that it does not have to reflect any particular ratio of hyperfine constants.  \\

\textbf{$^{199}$Hg, $\alpha$-line.} In Fig.~\ref{fig:Hg_WvsK} (a) we plot the linewidth against the shift for the Hg $\alpha$-line and $c\parallel{B_0}$. We find a dependence similar to that of Cu. There is again a linear regime (between about 80 K and 170 K), a high-$T$ drop, a low-$T$ upturn, as well as a $T$ independent part. With our model \eqref{eq:LinewidthX9702} we deduce from the slope that $\sqrt{\left< h^2 \right>}/\left<H\right>_{\rm X97} \approx$0.08 and thus a value that is less than half of what we found for Cu, while the drop of about 3.3~kHz or 0.0037\% is expected from the ratio of the hyperfine coefficients for component $C_X$. In order to understand this behavior we would like to point out that the Hg nucleus couples to \textit{two} Cu nuclei in the nearest two CuO$_2$ planes. Consequently, we have to replace $\sqrt{\left< h^2 \right>}$ by $\sqrt{\left[\left< h_i^2 \right>+\left<h_ih_j\right>\right]}/\sqrt{2}$, where $h_i$ and $h_j$ denote the effective charge variation in the CuO$_2$ plane above and below the Hg nucleus, respectively. Our results then say that as far as component $A$ is concerned, the correlation between the two planes appears to be negative. On the other hand, it appears to be close to 1 for the $C$ term. This says that the local deviation from $\left<H\right>$ is different for $A$ than for $C$, which points to different wavelength for both spin components. This is perhaps what one might expect if $A$ is a local property, contrary to $C$.

From the low-$T$ intercept in Fig.~\ref{fig:Hg_WvsK} we also deduce that there is a $T$ independent residual Hg NMR linewidth that could range between $^{199}\zeta_{\parallel,\alpha,\rm X97}=$ 8.5 kHz to 9.5 kHz for $c\parallel{B_0}$ (the uncertainty comes from the low-$T$ shift that could be influenced by diamagnetism). For $c\bot{B_0}$ (plot not shown) we estimate $^{199}\zeta_{\bot,\alpha,\rm X97} = 10$ kHz, and thus $^{199}\zeta_{\parallel,\alpha, \rm X97}/^{199}\zeta_{\bot,\alpha, \rm X97}\approx 0.9$. From the high-$T$ shifts we also deduce $^{199}{ p_\parallel}/^{199}{p_\bot} \approx 1.7$, which indicates that the residual broadening may not be dominated by a coupling through $^n{p}$ unless wavelength effects are of importance. On the other hand, $^{63}\zeta_{\parallel,\rm X97}/^{199}\zeta_{\parallel,\alpha, X97}\approx 11$ and thus similar to $^{63}{p_\parallel}/^{199}{p_\parallel}$, which suggests that the residual broadening $\zeta$ is not of orbital origin.

\begin{figure}
\center{}
\includegraphics[scale=0.75]{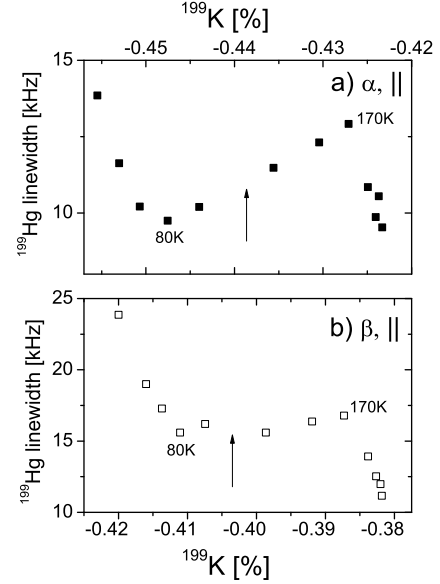}
\caption{$^{199}$Hg linewidth for the X97 crystal for $c\parallel B_0$ as a function of the shift, $^{199}K$ with temperature as an implicit parameter for a) $\alpha$ line and b) $\beta$ line. Arrows indicate $T_{\rm c}$.}
\label{fig:Hg_WvsK}
\end{figure}

The low-$T$ upturn (4~kHz) of the width from 80~K to 50~K for $c\parallel{B_0}$ would amount to a Cu linewidth of about 49~kHz for coupling through $^n{p}$, a value that appears far too big, cf. Fig.~\ref{fig:63LWK_X97}. On the other hand, if its origin is Meissner diamagnetism we estimate a 6~kHz increase in Cu linewidths in that $T$ range, which fits the experimental results. We conclude that the low-$T$ upturn is dominated by Meissner diamagnetism in the mixed state.

\textbf{Hg $\beta$-line.} In the lower panel of Fig.~\ref{fig:Hg_WvsK} we plot width vs. shift of the $\beta$-line for $c\parallel{B_0}$. The central, linear region has an even smaller slope of $\sqrt{\left< h^2 \right>}/\left<H\right>\lesssim0.05$ (it was 0.08 for the $\alpha$ site) indicating stronger negative correlations of the effective charge in adjacent planes. The high-$T$ drop of about 5 kHz or 0.0056\% appears at the same $T$ and is in agreement with the ratio of the hyperfine coefficients $^{199}{p}_\beta/^{199}{p}_\alpha \approx 1.4$.\cite{Haase2012} In addition, the $\beta$ site is close to the hole donating O$^{2+}_{\delta}$ and the average charge density $\left<H_\beta\right>$ might be higher and more homogeneous in its vicinity (the so-called Cu B-site in La$_{2-x}$Sr$_x$CuO$_4$ shows a similar phenomenon, i.e., a higher hole concentration that depends much less on doping, compared with the Cu A-site).\cite{Ohsugi1994, Haase2000a} 

The low-$T$ intercept of the width for vanishing shift $^{199}\zeta_{\parallel,\beta,X97} \approx 13$ kHz or 0.0145\% would fit to a spin component coupling to the $\beta$ Hg nuclei through $^{199}{p}_\beta$ (we do not have reliable data for $c\bot{B_0}$).
The low-$T$ upturn of about 8~kHz is somewhat in excess of what one might expect if the field distribution in the mixed state is the same for both sites, but since it is based on just one data point it is not significant.

From the above discussion or by inspecting again Fig.~\ref{fig:Hg_WvsK} it is clear that if we were to plot the linewidths of both Hg sites against each other, we would not obtain a straight line between 80 and 300 K. Indeed, in such a plot (not shown) coming from high $T$ we find that both widths increase with the expected slope of $^{199}{p}_\beta/^{199}{p}_\alpha \approx 1.4$ (with a low-$T$ intercept near [0,0] within limits of error). However, at about 170~K the slope sharply turns rather flat since now the $\beta$-site changes more slowly with $T$ compared to the $\alpha$-site. It takes another sharp turn towards slope of one below 80~K when the fluxoid lattice is established.

Interestingly, the temperature at which the linewidths have a maximum ($\sim$170-200 K) corresponds very well to the onset of the magnetic excitation recently found from neutron scattering\cite{Li2010} and the maximum of the thermoelectric power.\cite{Yamamoto2000} 
\begin{figure}
\center{}
\includegraphics[scale=0.75]{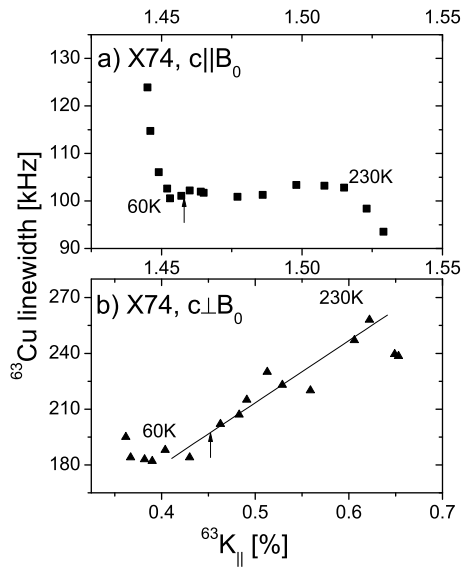}
\caption{$^{63}$Cu central transition linewidth for the X74 crystal as a function of the shift, $^{63}K$ with temperature as an implicit parameter for a) $c\parallel B_0$ and b) $c\perp B_0$. The straight line is a fit and the arrows indicate $T_{\rm c}$.}
\label{fig:63LWK_X74}
\end{figure}
\subsubsection{X74 Linewidths}
For the X74 crystal the $^{63}$Cu linewidth vs.~shift plots in Fig.~\ref{fig:63LWK_X74} show a high-$T$ drop now at 230 K and low-$T$ upturn also at about 60 K. However, while the linewidth between 60 K and 230 K increases for $c\bot{B_0}$, it is quite flat in $T$ for ${c\parallel{B_0}}$. The $T$ independent width for ${c\parallel{B_0}}$ is about $^{63}\zeta_{\parallel,X74} \approx$ 90 kHz, similar to what we found for X97. For $c\bot{B_0}$ we find a $T$ independent width of approximately 170 kHz that includes the quadrupolar broadening of about 140 kHz ($\lambda_{\rm Q,X74}\approx 2.5$ MHz/hole). The additional magnetic broadening for $c\bot{B_0}$ could thus range from 30 kHz to about 100  kHz for coherent and incoherent broadening, respectively. Near $T_{\rm c}$, we do not observe substantial changes in the linewidths for either orientation. 
The high-$T$ downturn is about 1/3 of that found for X97. The low-$T$ upturn agrees with the findings for X97.

Most surprising is the fact that between 60 K and 230 K there is no $T$-dependent linewidth for ${c\parallel{B_0}}$ since we expect $A_{\rm X74}$ to vary substantially across the CuO$_2$ plane similarly to $A_{\rm X97}$ ($\sqrt{\left< h^2 \right>}$ is the same for both crystals). In fact, in our model (Eq. \ref{eq:LinewidthEqn}) the change of the linewidth between 60 K and 230 K should amount to {$\Delta K_{X74, \parallel} \cdot \sqrt{\left< h^2 \right>}/\left<H\right>_{X74}\approx$ 20 kHz, which is clearly in excess of what we find experimentally in this temperature range (Fig. \ref{fig:63LWK_X74}a). Note that our model (Eq. \ref{eq:LinewidthEqn}) does not contain the additional $T$ independent broadening $\zeta$ that we also found for X97. If this broadening $^{63}\zeta_{\parallel,X74} \approx$ 90 kHz were incoherent with regard to that caused by $A_{\rm X74}$, the linear dependence would indeed almost disappear. However, since the broadening through $A_{\rm X74}$ must be coherent with the quadrupole splitting that was calculated to cause a 140 kHz linewidth for $c\bot{B_0}$, the broadening from A$_{\rm X74}$ should be lifted out of the incoherent $\zeta$ broadening by this second-order quadrupole term for $c\bot{B_0}$. This is indeed the case as Fig.~\ref{fig:63LWK_X74}b reveals. Between 60 K and 230 K the linewidth changes by $\approx$75 kHz, in agreement with what we expect from our model. 

The Hg NMR linewidth versus shift plot for ${c\parallel{B_0}}$ (not shown) is flat in $T$ between about 60 K and 200 K and has a high-$T$ downturn, as well as a low-$T$ upturn, all in agreement with the Cu data.

For the X74 crystal the temperature at which the linewidths have a maximum ($\sim$230 K) also corresponds very well with temperature at which a maximum of the thermoelectric power was observed\cite{Yamamoto2000} and with $T^{**}$ (when $\rho$ becomes $\propto T^2$).\cite{Barisic2012} For the X74 crystal we measured only up to 300 K, i.e. below the onset of the novel magnetic excitation $T^{*}$.   

\subsubsection{Linewidths in the mixed state}
While we could not investigate the low-$T$ broadening down to the lowest $T$ with our single crystals, we believe that the increase of linewidths at low temperatures is due to vortex lattice, which forms in the mixed state.\cite{Pincus1964, Brandt1988} The broadening of NMR lines starts at 60-70 K, which is somewhat lower than T$_{\rm c}$, but it agrees with the recent SANS results on nearly optimally doped HgBa$_{2}$CuO$_{4+\delta}$, in which the vortex lattice signal could be observed only below 80 K.\cite{Li2011} In the vortex phase the local magnetic field is spatially distributed resulting in an increase of the NMR linewidth. A similar low temperature behavior of the linewidth was observed for powder samples\cite{Itoh1998} and for other cuprates, as well. \cite{Corti1996, Curro2000} 

We can estimate effects of this broadening, assuming an ideal 3D triangular vortex lattice for the $c \parallel B_0$ orientation, with $B_{\rm c1}<<B_0<<B_{\rm c2}$, the distribution of the local field is given by: \mbox{$0.0609 \phi_0 / \lambda_{\rm ab}^2$}, \cite{Brandt1988} where $B_{\rm c1}$ and $B_{\rm c2}$ are lower and upper critical fields, respectively, $\phi_0 = 2.07\times10^{-15}$ Tm$^2$ is the magnetic flux quantum and $\lambda_{\rm ab}$ is the in-plane penetration depth. For the optimally doped crystal $\lambda_{\rm ab}(T=0$K)$ \cong 150$ nm.\cite{Lebras1996, Hofer1998, Villard1999} Thus, at T=0 K we expect a broadening of about 60 kHz and 40 kHz for $^{63}$Cu and $^{199}$Hg, respectively. For the X74 crystal $\lambda_{\rm ab}(0)$ increases to about 235 nm\cite{Villard1999} and we can expect additional broadening of 25 kHz for $^{63}$Cu and 17 kHz for $^{199}$Hg. 

Our experimental resonance lines have a Gaussian shape, not the shape characteristic for a field distribution in the vortex phase,\cite{Brandt1988} which is obscured by already large magnetic widths. 
In general the linewidth due to the vortex lattice could be coherent or incoherent with $\zeta$. Using $^{63}$Cu data, for which we have measured down to 20 K, we can test which case agrees better with theoretical values. For the X97 crystal we estimate a vortex lattice contribution of 50 kHz and 120 kHz for the coherent and incoherent case, respectively. Clearly the coherent case agrees better with theory (60 kHz). The same conclusion is made for the X74 crystal, we obtain 30 kHz (80 kHz) for the coherent (incoherent) case and the theoretical value is 25 kHz. 
Assuming the coherent case for $^{199}$Hg, the linewidth increase due to vortex lattice is up to 10 kHz (4 kHz) for the X97 (X74) crystal. These values are smaller than expected, but we measured $^{199}$Hg only down to 50 K (30 K).
For $c \perp B_0$ all linewidths show similar low temperature behavior. Although we did not measure at temperatures close to 0 K, our experimental values are in reasonable agreement with theoretical predictions. 

\section{Conclusions}
We have presented extensive measurements of the $^{63}$Cu quadrupole parameters, shifts, and linewidths, as well as the shifts and linewidths of $^{199}$Hg for two single crystals of HgBa$_{2}$CuO$_{4+\delta}$, one underdoped with $T_{\rm c}$=74 K and one optimally-doped with $T_{\rm c}$=97 K. The $^{63}$Cu quadrupole interaction, which is largely $T$ independent, is in agreement with earlier studies on powder samples. We determine that the asymmetry parameter is very small ($\eta < 0.006$), despite a rather large variation of the quadrupole splittings that we believe is caused by a charge density variation in the CuO$_2$ plane (as found in other non-stoichiometric HTSC). 

With the knowledge of the quadrupole interaction we can measure the Cu magnetic shifts and we find that they cannot be explained with a single $T$-dependent spin component (the same conclusion is reached just from the Hg magnetic shifts, as well as from comparing the Cu and Hg magnetic shifts\cite{Haase2012}). We find that we can explain all the shifts with \textit{two} spin components that have different $T$ dependences. The $T$ dependence of the first component carries the pseudogap behavior, i.e., it is $T$-dependent already above $T_{\rm c}$ and does not change at $T_{\rm c}$. The other component is $T$ independent above $T_{\rm c}$ similar to the behavior of a Fermi liquid and vanishes rapidly below $T_{\rm c}$. The susceptibilities of both components grow with doping at a given $T$. 

With $^{199}$Hg NMR we can resolve two resonance lines (that could not be distinguished before) with different, but symmetric chemical shift tensors. One Hg site is neighbored by doped extra oxygen O$_\delta$. The fact that only one Hg site appears as a result of doping points to a well defined chemical structure, despite the electronic inhomogeneity in the CuO$_2$ plane. The intensity ratio of the lines is in agreement with what one expects from interstitial oxygen in the middle of an Hg plaquette that changes the chemical shift of the four nearest Hg atoms. The Hg magnetic shifts also reveal two-component behavior and we can estimate the Meissner diamagnetism from the shifts at low $T$.

The study of the Cu satellite and central transition linewidths, as well as the magnetic linewidths of both Hg resonances reveal the following simple features:

(1) The Cu satellite widths are largely $T$ independent due to a $T$ independent charge density modulation in the CuO$_2$ plane. 

(2) Below about 60 K the magnetic widths for Cu and Hg are dominated by the fluxoid lattice (which was not studied in great detail here due to penetration depth issues for single crystals at very low $T$). 

(3) Above 60 K all magnetic widths can be understood in terms of spatial shift variations (long wavelengths) from the two spin components, in addition to a $T$ independent broadening due to an unknown spin component. Above about 200 K to 230 K the Fermi-liquid-like component becomes suddenly homogeneous (its contribution to the width disappears), while that due to the pseudogap component remains. Interestingly, all these linewidths mechanisms are coherent with the charge density modulation in the CuO$_2$ plane for the optimally doped sample, but for the underdoped sample only the $T$ independent component becomes incoherent.\\

The two spin components that give rise to the $T$-dependent magnetic shifts are in agreement with what has been found in La$_{2-x}$Sr$_x$CuO$_4$ and YBa$_2$Cu$_4$O$_8$.\cite{Haase2009, Meissner2011} The fact that we confirm them with shift measurements on a very different system and by comparing different nuclei\cite{Haase2012} makes them ubiquitous to the HTSC. Only with a more detailed investigation of the two components can we point clearly to their nature, however, they do resemble the susceptibilities of a spin liquid and that of a Fermi liquid. The universal presence of a Fermi-liquid-like component has recently been established from other experiments as well.\cite{Barisic2012, Mirzaei2012}

The results of our linewidths studies appear to be in agreement with the observation that large electronic inhomogeneities can be present in HTSC without degrading $T_{\rm c}$. In particular, our studies of the bulk of the samples appear to be in agreement with surface studies that find the pseudogap to be rather inhomogeneous,\cite{Boyer2007, Pushp2009} since our shift variations are dominated by the pseudogap component. We note that stoichiometric compounds do not have this static inhomogeneity. The onset of motional narrowing or depinning at 170-200 K (X97), 230 K (X74) is unexplained. However, these temperatures correspond very well to $T^{**}$, a temperature at which a maximum of thermoelectric power and onset of $\rho \propto T^2$ (a Fermi-liquid-like behavior) were observed.\cite{Yamamoto2000, Barisic2012} Additionally, for crystals close to optimal doping a novel magnetic excitation appears at around 200 K.\cite{Li2008a, Li2010, Li2011a, Li2012} It is unclear, whether the coherent modulation of the charge density with the various shift components could have to do with a "stripy" electronic state that may exhibit such features in the NMR when pinned to the lattice.

\section{Acknowledgments}
We acknowledge the financial support from the University of Leipzig and the DFG within the Graduate School Build-MoNa and ESF project no. 080939247. We also acknowledge the assistance of T. Meissner and M. Bertmer and discussions with C. P. Slichter and O. P. Sushkov (Leipzig group). The crystal growth and characterization work was supported by the US Department of Energy, Office of basic Energy Sciences.  X.Z. acknowledges support by the National Natural Science Foundation of China (Grant No. 20871052).

\bibliography{HTSC}

\end{document}